\documentclass[aps,amsfonts,nofootinbib,superscriptaddress,twocolumn]{revtex4-1}

\usepackage{appendix}
\usepackage{dsfont}
\usepackage{amssymb}
\usepackage{amsmath}
\usepackage{amsbsy}
\usepackage{color}
\usepackage{slashed}
\usepackage{bm}
\usepackage{bbm}
\usepackage{dutchcal}

\newcommand{\ee}{\end{equation}}
\newcommand{\bb}{\begin{equation}}
\newcommand{\eqb}{\begin{eqnarray}}
\newcommand{\eqf}{\end{eqnarray}}

\begin{document}
\title{ Classical Noncommutative Bicosmology Model}
\author{H. Falomir }
\email{falomir@fisica.unlp.edu.ar}
\affiliation{Departamento de  F\'{\i}sica, Universidad Nacional  de La
  Plata, La Plata, Argentina}
 \author{J. Gamboa }
\email{jorge.gamboa@usach.cl}
\affiliation{Departamento de  F\'{\i}sica, Universidad de  Santiago de
  Chile, Casilla 307, Santiago, Chile}
  \author{F. M\'endez }
\email{fernando.mendez@usach.cl}
\affiliation{Departamento de  F\'{\i}sica, Universidad de  Santiago de
  Chile, Casilla 307, Santiago, Chile}
\date{\today}

\begin{abstract}
  We  propose a  bicosmology model  which is  the classical  analog of
  noncommutative  quantum  mechanics.  From  this point  of  view  the
  sources of the modified FRW  equations are dark energy ones governed
  by a Chapligyn's equation  state. The parameters of noncommutativity
  $\theta$ and $B$  are interpreted in terms of the  Planck area and a
  like-magnetic field, presumably the magnetic seed of magnetogenesis.
\end{abstract}
\date{\today}
\maketitle


\section{ Introduction}

In  the last  forty years  a lot  of observational  evidence has  been
accumulated  showing   a  remarkable   agreement  with   the  standard
cosmological model.  These observations also show that the universe is
expanding  at an  accelerated rate  and  then, the  model requires  to
incorporate  dark  energy  in   order  to  explain  such  acceleration
\cite{trod}.

The mere existence  of dark energy requires ideas  beyond the standard
cosmological  model, in  a  similar  way that  the  standard model  of
particles needs to be modified in order to incorporate the dark matter
\cite{some,recent,varios,planck}.

However, the  incorporation of  dark energy might  shed some  light on
other long-standing problems in cosmology.   One of such issues is the
origin  of  the magnetic  fields  in  galaxies  and the  mechanism  to
originate  a  magnetic seed  in  the  universe. These  magnetic  seeds
requires,  in principle  the break  of a  symmetry, which  is possibly
hidden, appearing  as an  effective degree of  freedom. This  would be
similar to the magnetic field of  a magnet, which has a purely quantum
origin.

From the cosmological point of view,  the scale factor in the
Friedmann-Robertson-Walker (FRW) solution describes
our universe as a bubble that evolves according to the FRW equations.

The assumed large scale homogeneity of the Universe leading to the FRW
solution also  implies that, seeing  from each point of  spacetime, it
evolves  as  a patch  causally  disconnected  from  the rest  of  the
Universe.  Therefore,  the  assumptions  of  the  existence  of  other
universes causally disconnected can not be theoretically ruled out.

The possibility that these  patches evolve and interchange information
between them would  require that the cosmological principle  be just a
\emph{small distance} approximation, otherwise  the formation of these
structures would not be entangled.

In this sense, if the cosmological principle is not an exact symmetry,
the transference  of information between  patches should be  a process
that would leave some observable traces. This poses the question about
the information transfer mechanism.

One of the goals of this letter  is to offer an approach that combines
the  idea of  bimetric gravity  \cite{bimetricgrav,g1,g2,g3,g4} and  a
relationship  with noncommutative  quantum  mechanics  (for a  general
discussion   on  bimetric   gravity  as   cosmology  see   {\it  e.g.}
\cite{del}).

The  idea sketched  above allows  to incorporate  considerations about
gauge invariance, causality and evolution  of states, thus providing a
different approach to those modern cosmological problems.

The approach has interesting  implications because allows to
give a physical interpretation to  the infrared and ultraviolet scales.

The letter  is organized  as follows:  In section  II we  will shortly
review some basic aspects of the  FRW metric in order to fix notations
and to  establish some basic features  which will be necessary  to our
model.

Sections III and  IV are devoted to present and  develop the model. The
last section V contains the final remarks.

\section{basics issues}

Let us start  discussing the notation: Firstly we consider
the metric
\bb
ds^2 =-N(t)^2\,dt^2 + a(t)^2 \left(\frac{dr^2}{1-kr^2} 
  + r^2d\Omega^2\right), 
\label{RW}
\ee 
where coordinates  are $\{t,r,\theta,\varphi\}$, $k=\{-1,0,1\}$ is
the spatial curvature and $N(t)$ is the lapse shift.

The  equations of  motion for  metric  (\ref{RW}) reduce  to the  FLRW
equations which, in the gauge $N=1$, turn out to be
\eqb
\label{frw1}
2\frac{\ddot  a}{a}  +\left(\frac{\dot   a}{a}\right)^2  &=&  \Lambda+
\cdots,
\\
\left(\frac{\dot a}{a}\right)^2 &=& \frac{1}{3}\Lambda+ \cdots,
\label{frw2}
\eqf
where $\cdots$ denote contributions from matter fields. From here on we restrict
to the case $k=0$, which  is also consistent with present observations
\cite{pdg} and, for the sake of  clarity, we will also omit the matter
contributions. 

 The previous equations can also be obtained from the Lagrangian
\bb
L= \frac{1}{2N} a {\dot a}^2 + \frac{N}{6}\Lambda a^3.
\label{a1}
\ee
Indeed, while equation (\ref{frw1}) results from variations with respect to
the variable $a$, the second  one (\ref{frw2}) results from variations
respect to $N$. 

Consider now  the following change of variables
\bb
x=\frac{2}{3\sqrt{G}} a^{3/2}.
\label{cambio}
\ee
In   terms   of  this   new   variable,   which  has   dimensions   of
(energy)$^{-1/2}$, the Lagrangian in (\ref{a1}) changes into
\bb
G^{3/2}\,L= \frac{1}{2N} {\dot x}^2 - \frac{N}{2} \omega^2 x^2,
\label{2b}
\ee
where the frequency is defined as
$$
\omega^2 = -\frac{3}{4}\Lambda.
$$
From here on we will omit the global factor $G$ in the Lagrangian. 

The equations  of motion obtained  though variations of  the variables
$x$ and $N$ are 
\eqb
{\ddot x} + \omega^2 x&=&0,
\label{3aa}
\\
{\dot x}^2 + \omega^2 x^2&=&0.
\label{3a}
\eqf
Note that the constraint (\ref{3a}) is obtained also from (\ref{3aa}) as
a first integral. Indeed, multiplying this last equation by $\dot{x}$ we
get the equivalent equation 
\[
\frac{d}{dt} \left[ {\dot x}^2 + \omega^2 x^2\right]=0.
\]
Therefore,  the  constant  ${\dot x}^2 + \omega^2 x^2$ must be chosen
equal to zero (in which case the
energy is zero) and non trivial solutions are obtained for $\omega$ an
imaginary  number,  as   it  is  upon  the   identification  with  the
cosmological constant. 

\section{Modified FRW equations}
In this section we will modify the FRW equations by assuming more than
one scale factor.   This could happen, for example,  in bubbles models
(these bubbles have been extensively studied in recent literature, see
{\it e.g.} \cite{kleban}).

The  assumptions of  homogeneity  and isotropy  in  the present  model
implies that these scale factors $ a_i (t)$, $i=1, 2, \cdots$ satisfy 
\bb
\left[ a_i (t), a_j(t)\right]=0, \label{micro1}
\ee
where $[~,~]$ denotes a Poisson bracket.

Note  that this  \lq \lq  microcausality\rq \rq  ~principle implies  the
possibility of  choosing only one  lapse function and,  therefore, the
existence of {\it one} cosmological time. 

The  key  observation is  to  note  that  in the  harmonic  oscillator
representation  (\ref{cambio}), the  model  with more  than one  scale
factor admits a straightforward generalization.

Indeed, instead  of the  usual Poisson's  bracket of  momenta $p_i(t)$
satisfying   $[p_i   (t),    p_j   (t)]   =   0$,    we   can   choose
$[p_i (t),  p_j (t)]  = F_{ij}$  where $F_{ij}$  is a  constant tensor
which can  be identified  with some internal  degree of  freedom which
acts as an  effective infrared cut-off. From now on,  we will consider
just two patches and write  $F_{ij}= \epsilon_{ijk}B_k$ for a constant
$B$.

In the harmonic oscillator representation of (\ref{2b}), we can choose
the Lagrangian 
\bb 
L_0= \frac{1}{2N} \left({\dot x}_1^2+{\dot x}_2^2\right) - \frac{N}{2}
\left(\omega_1^2 x_1^2+ \omega_2^2 x_2^2\right), 
\label{la11}
\ee
 which  is the  generalization  of (\ref{2b})  which incorporates  two
 scale factors with frequencies 
\bb
\omega_1^2=   -\frac{3}{4}   \Lambda_1,   ~~~~~~~~~~~~~~   \omega_2^2=
-\frac{3}{4} \Lambda_2.
\label{cosmo22}
\ee

We also add an interaction term
\bb
{\bar L}= \frac{B}{2}\left(x_1 {\dot x}_2 -x_2{\dot x}_1\right)
\label{a4}
\ee
to get the total Lagrangian
\bb
L = L_0 + {\bar L}. \label{a001}
\ee

This Lagrangian formally describes  a charged nonrelativistic particle
moving  in  a  constant  magnetic   field  $B$  perpendicular  to  the
$\langle x_1,  x_2 \rangle$  plane, {pointing in  the direction  of an
  \emph{$x_3$-axis}.   Note   that   one  could   define   the   ratio
  ${B}=\frac{B}{\sqrt{G}}$   with  dimensions   (energy)$^{+1}$  where
  $\frac{1}{\sqrt{G}}$ is the Plack mass.

The equations of motion (in the gauge $N=1$) are
\eqb
&&{\ddot x}_1 +\omega_1^2~ x_1  -{\cal B}~ {\dot x}_2=0,
\nonumber
\\
&&{\ddot x}_2 +\omega_2^2 ~x_2 +{\cal B} ~{\dot x}_1=0,
\label{i02}
\eqf
while the constraint, a   consequence  of   time
reparametrization invariance, reads as
\bb
\frac{1}{2}   \left(   {\dot  x}_1^2+   {\dot   x}_2^2\right)   +
\frac{1}{2}\left(\omega_1^2 x_1^2+\omega_2^2 x_2^2\right)=0.
\label{i03}
\ee

As in the previous section, the condition (\ref{i03}) gives nontrivial
solutions for imaginary frequencies, which is just the case at hand.

In  order  to  make  contact with  the  cosmological  description  and
following the analogy with (\ref{cambio}), if we define 
\bb
x_1 = \frac{2}{3\sqrt{G}} a^{3/2}, ~~~~~~x_2 = \frac{2}{3\sqrt{G}}b^{3/2},
\label{cambio2}
\ee
equations (\ref{i02}) become
\eqb
 2\,  \frac{{\ddot  a}}{a}  +  \left(\frac{{\dot  a}}{a}\right)^2  &=&
 -\frac{4}{3}\omega_1^2 +  {\mathcal B} \sqrt{a\,b}~ \frac{{\dot b}}{a^2},
\nonumber
\\
 2\, \frac{{\ddot b}}{b}+ \left(\frac{{\dot b}}{b}\right)^2 &=&
-\frac{4}{3}\omega_2^2 - {\mathcal B} \sqrt{a\,b} \frac{ {\dot a}}{b^2},
\label{i220}
\eqf
and the constraint  (\ref{i03}) now reads as
\bb
\label{i230}
\left(\frac{ {\dot  a}}{a}\right)^2 =
-        \left(\frac{2}{3}\omega_1\right)^2-       \frac{1}{a^3}\left(
  \frac{3}{4}\omega_2^2 b^3 +{\dot b}^2 b\right). 
\ee

The expressions in (\ref{i220}) can be considered as the FRW equations
for two patches of the Universe that interact through a \emph{constant
  external like-magnetic  field} while  (\ref{i230}) is the  analog of
$G_{00}= -8\pi G ~ T_{00}$ in the one-metric conventional gravity.

We would like to emphasize that,  in this picture, dark energy emerges
as a consequence of the incorporation of a sort of interaction between
neighboring patches in spacetime.

From the point of view of the energy and momentum content in sector 1,
eqs.\ (\ref {i220}) and (\ref {i230}) turn out to be
$$
T_{11}=T_{22}=T_{33}= - {\mathcal B} \sqrt{a\,b}~ \frac{{\dot b}}{a^2},
$$ 
and
$$
T_{00}=    -\left(\frac{b}{a}\right)^3     \left[    \left(\frac{{\dot
        b}}{b}\right)^2  +\left(\frac{2}{3}\omega_2   \right)^2  \right]
-\left(\frac{2}{3}\omega_1 \right)^2.
$$

From these results one has the following state equation,
\bb
\rho_b + \frac{6\pi}{{\mathcal B}^2}  p_b^2 = \frac{\Lambda_2}{8\pi G}
\left(\frac{b}{a}\right)^3, 
\ee
which describes a Chapligyn gas. This gas has been discussed extensively
in   cosmology   in   connection   with  the   dark   energy   problem
\cite{gorini,trod1}. 

It is also interesting to discuss  this problem from the point of view
of a  Hamiltonian system with  modified Poisson Brackets.  Indeed, the
mechanical system described by the Lagrangian in (\ref{a001}) can also
be described by the Hamiltonian
 \bb
   H = \frac{N}{2} \bigg(p_1^2+p_2^2  +
\omega_1^2x_1^2+ \omega_2^2x_2^2 \bigg),
\label{elha}
\ee
with the Poisson's bracket algebra
\begin{equation}
\label{15}
\left[x_i,x_j\right] = 0,\quad \left[x_i,p_j\right] = \delta_{ij},
\quad
\left[p_i,p_j\right]  = \epsilon_{ij} \, B.
\end{equation}
Note that  $x$ has dimensions  of (energy)$^{-1/2}$ while $p$  has its
inverse  dimensions. Then  $B$ has  dimensions of  (energy)$^{+1}$ and
therefore, the magnetic field is $B/\sqrt{G}$.

\section{Noncommutative Classical Cosmology}

We  observe that  it  is still  possible to  consider  a more  general
deformation of the  Poisson's algebra, that is, the  introduction of a
noncommutative parameter in the bracket coordinates.

With this  deformation (as  in \cite{np}) we  have two  energy scales,
namely the Planck  energy $E_P=G^{-\frac12}$ and the  \lq \lq magnetic
energy"  $B^{\frac12}$  (or,  equivalently, the  Planck  and  magnetic
lengths), since  $G$ (the  Newton constant)  and $B$  (a magnetic-like
seed)  define two  natural  scales for  the  modified Poisson  bracket
structure.

In order to explore this system we restrict ourselves to the case with
$\omega_1=\omega_2\equiv\omega$ and rewrite  (\ref{elha}) as
\bb
H = N {\cal H},
\ee
where the constraint reads now as
\bb
{\cal  H}=  \frac{\omega}{2} \left({\bar  p}_1^2+ {\bar p}_2^2  + {\bar
    x}_1^2 +{\bar x}_2^2\right),
 \label{elha1}
\ee
with  original    phase space variables  $\{x_i,p_j\}$     rescaled
according to    $x_i  \to \bar{x}_i = \sqrt{\omega} x_i={\bar x}$
and $p_j \to \bar{p}_j= p_j/\sqrt{\omega}$.

The modified Poisson brackets structure, in view of
our previous discussion   \cite{np,bellu,ccgm}, is
\eqb
 \left[{\bar x}_i,{\bar x}_j\right] &=&\epsilon_{ij}\,{\bar G}, 
 \label{150}
 \\
 \left[{\bar x}_i,{\bar p}_j\right] &=& \delta_{ij},
 \label{150B}
 \\
 \left[{\bar p}_i,{\bar p}_j\right] &=& \epsilon_{ij}\,{\bar B},   
 \label{151}
 \eqf
 where $\bar{G}$ and $\bar{B}$ are  the rescaled Newton's constant and
 magnetic-like       seed,       respectively,      according       to
 $          {\bar          G}=          \sqrt{G}\,\omega$          and
 ${\bar          B}=          \frac{B}{\omega}          =\sqrt{G}{\cal
   B}_{\mbox{\tiny{seed}}}/\omega                $                with
 ${\cal B}_{\mbox{\tiny{seed}}} = B/\sqrt{G}$

The equations  of motion for the variables $\bar{x}_i$ turn out to be
\bb
\ddot{   \bar{    x}}_i   +   {   \Omega}^2    \,\bar{x}_i   -   {\cal
  B}\,\epsilon_{ij}\, \dot{\bar{ x}}_j = 0.
\label{eq12}
\ee
where
\eqb
\Omega^2   &=&   \omega^2(1   -   {\bar   G}{\bar   B})   =   \omega^2
\left(1-GB_{\mbox{\tiny{seed}}} \right),
\label{t1}
\\
{\cal    B}   &=&\omega(    {\bar   B}+    {\bar   G})    =   \sqrt{G}
\left(B_{\mbox{\tiny{seed}}} + \omega^2\right).
 \label{t2}
\eqf

Comparing    equations    (\ref{eq12})    with    (\ref{i02})    (with
$\omega_1=\omega_2=\omega$ for the last case) we see that an effective
magnetic background field, given by
\begin{equation}
\label{effb}
{\cal  B}/\sqrt{G} = B_{\mbox{\tiny{seed}}} +\omega^2,
\end{equation}
 is generated.

 We identify ${\cal B}/\sqrt{G}$ as the effective magnetic seed in the
 universe,  ${\cal B}_{\mbox{\tiny{eff}}}$,  while  $\Omega^2$ can  be
 seen as an effective cosmological constant.

 Note    that,     as    in    noncommutative     quantum    mechanics
 \cite{np,bellu,ccgm}, there are  two phases with $G {\cal  B} >1$ and
 $G  {\cal B}  <1$  respectively,  separated by  a  critical point  at
 $G {\cal B}= 1$.

If we demand $GB_{{\mbox{\tiny{seed}}}} \ll 1$, then
$$
B_{{\mbox{\tiny{seed}}}} \approx {\cal  B}_{\mbox{\tiny{eff}}}+\frac{3}{4}\Lambda.
$$

This  formula   provides  a  simple   and  direct  link   between  the
magnetogenesis  \cite{holland,ratra}  and  the  cosmological  constant
problem.

{Note that the modified  Poisson brackets in (\ref{150}), (\ref{150B})
  and  (\ref{151})  can  be  mapped   to  a  canonical  form  under  a
  non-canonical             change            of             variables
  $\{\bar{x}_i,\bar{p}_j\}\to\{X_i,P_j\}$ where the new Hamiltonian is
  still diagonal \cite{np,ccgm}.   Thus, ${\cal H}$ can  be written as
  \bb
\label{Hdiago}
\frac{H}{N}  = \frac{1}{2}\left(  P_1^2  +  \Omega_+^2 X^2_1\right)  +
\frac{1}{2} \left(P_2^2 + \Omega_-^2 X_2^2\right), 
\ee 
where  variables  $\{X_1,P_1\}$   and  $\{X_2,P_2\}$  are  canonically
conjugated, while  the variables  of the sector  labeled as  \lq1\rq ~
have  zero  Poisson  bracket  with   those  of  sector  \lq2\rq.   The
frequencies in (\ref{Hdiago}) are \cite{np,ccgm}
\begin{eqnarray}
\Omega_\pm &=& \pm \omega \left[\sqrt{1 + \frac{1}{4} ({\bar B}- {\bar
      G})^2} \mp \frac{1}{2}\left({\bar B}+ {\bar G} \right)\right],
      \\
      &\approx&\pm\left[\sqrt{\omega^{2} +\left(\frac{ \sqrt{G}\,{\cal
                B}_{\mbox{\tiny{eff}}}}{2}\right)^{2}} \mp 
     \frac{ \sqrt{G}\,{\cal  B}_{\mbox{\tiny{eff}}}}{2}\right],
\end{eqnarray}
where,    in   the    last    line,   we    used    the   fact    that
$B_{\mbox{\tiny{seed}}} - \omega^{2}  = {\cal B}_{\mbox{\tiny{eff}}} -
2\omega^{2}  \approx  {\cal B}_{\mbox{\tiny{eff}}}$  when  $\omega^{2}
\approx -  3 \Lambda/4  = \Omega^{2}$.   With same  approximations, we
finally obtain $\Omega_{\pm} ^{2}\approx \omega^{2}$.}

\section{Final Remarks}

We should also  observe that there are
  differences  between  the  quantum  Hall  effect  approach  and  the
  magnetogenesis as discussed in \cite{holland,ratra,us2}.  Indeed, as
  argued in  \cite{us2}, the  causal connection between  two spacetime
  regions  take  place  in  the   presence  of  an  external  constant
  magnetic-like field (at first sight presumably from the formation of
  some  galactic halo).   This external  magnetic-like field  would be
  very small, but not necessarily  a magnetic seed.  The magnetic seed
  would be  created much  earlier, as  can be  seen from  the magnetic
  displacement (\ref{t2}).

However,  this posses  some intriguing
  questions:  Is  ${\cal B}$  a  real  magnetic  field?, What  is  the
  mechanism  responsible for  the creation  of this  ${\cal B}$?.   At
  first glance,  it is interesting  to think  that the origin  of this
  field could be  similar to that of the magnetic  field produced by a
  magnet and, therefore, of purely quantum origin.

This work  was partially  supported by Dicyt-USACH.  H.F.  thanks ANPCyT, CONICET  
and  UNLP, Argentina,  for  partial  support through  grants PICT-2014-2304,  
PIP 688  and  Proy. Nro.  11/X748, respectively.  We would  like to
thank   M.~Paranjape    and   A.~P.~Polychronakos    for  enlightening
discussions.



\begin{thebibliography}{99}
\bibitem{trod} A.~Joyce, B.~Jain, J.~Khoury and M.~Trodden,
  Phys.\ Rept.\  {\bf 568}, 1 (2015)
 \bibitem{some} For a nice recent review see,  I.~Debono and G.~F.~Smoot,
  Universe {\bf 2}, no. 4, 23 (2016).




 \bibitem{recent} J.~Frieman, M.~Turner and D.~Huterer,
  Ann.\ Rev.\ Astron.\ Astrophys.\ {\bf 46} (2008) 385.




  \bibitem{varios} D.~F.~Mota and D.~J.~Shaw,
  Phys.\ Rev.\ D {\bf 75}, 063501 (2007);  F.~K.~Hansen, A.~J.~Banday and K.~M.~Gorski,
  Mon.\ Not.\ Roy.\ Astron.\ Soc.\  {\bf 354}, 641 (2004);   R.~Bousso, R.~Harnik, G.~D.~Kribs and G.~Perez,
  Phys.\ Rev.\ D {\bf 76}, 043513 (2007).



 \bibitem{planck} For example, P.~A.~R.~Ade {\it et al.} [Planck Collaboration],
  Astron.\ Astrophys.\  {\bf 594}, A20 (2016); P.~A.~R.~Ade {\it et al.} [BICEP2 and Planck Collaborations],
  Phys.\ Rev.\ Lett.\  {\bf 114}, 101301 (2015).


  \bibitem{bimetricgrav} Y.~Akrami, S.~F.~Hassan, F.~K\"onnig, A.~Schmidt-May and A.~R.~Solomon,
  Phys.\ Lett.\ B {\bf 748}, 37 (2015).
  \bibitem{g1}    Y.~Akrami, T.~S.~Koivisto, D.~F.~Mota and M.~Sandstad,
  JCAP {\bf 1310}, 046 (2013).
  \bibitem{g2} G.~Cusin, R.~Durrer, P.~Guarato and M.~Motta,
  JCAP {\bf 1505}, no. 05, 030 (2015).
  \bibitem{g3} S.~Deser, K.~Izumi, Y.~C.~Ong and A.~Waldron,
  Mod.\ Phys.\ Lett.\ A {\bf 30}, 1540006 (2015).
  \bibitem{g4} C.~de Rham, L.~Heisenberg and R.~H.~Ribeiro,
  Class.\ Quant.\ Grav.\  {\bf 32}, 035022 (2015)

  \bibitem{del}   K.~Bamba, A.~N.~Makarenko, A.~N.~Myagky, S.~Nojiri and S.~D.~Odintsov,
  JCAP {\bf 1401}, 008 (2014) and references therein.




  \bibitem{pdg}C. Patrignani et al. (Particle Data Group), Chin. Phys. C, {\bf40}, 100001 (2016) and 2017 update.



  \bibitem{kleban}
  M.~Kleban,
  Class.\ Quant.\ Grav.\  {\bf 28}, 204008 (2011); R.~Gobbetti and M.~Kleban,
  JCAP {\bf 1205}, 025 (2012); J.~L.~Lehners,
  Class.\ Quant.\ Grav.\  {\bf 28}, 204004 (2011);  A.~Aguirre, M.~C.~Johnson and A.~Shomer,
  Phys.\ Rev.\ D {\bf 76}, 063509 (2007); S.~Chang, M.~Kleban and T.~S.~Levi,
  JCAP {\bf 0804}, 034 (2008).
\bibitem{gorini} V.~Gorini, A.~Kamenshchik and U.~Moschella,
  Phys.\ Rev.\ D {\bf 67}, 063509 (2003);   V.~Gorini, A.~Kamenshchik, U.~Moschella and V.~Pasquier,
  gr-qc/0403062;   V.~Gorini, A.~Y.~Kamenshchik, U.~Moschella, O.~F.~Piattella and A.~A.~Starobinsky,
  JCAP {\bf 0802}, 016 (2008).
  \bibitem{trod1} A.~Melchiorri, L.~Mersini-Houghton, C.~J.~Odman and M.~Trodden,
  Phys.\ Rev.\ D {\bf 68}, 043509 (2003).
  
  \bibitem{holland} A.~Kandus, K.~E.~Kunze and C.~G.~Tsagas,
  Phys.\ Rept.\  {\bf 505}, 1 (2011).



  \bibitem{ratra} B.~Ratra,
  Astrophys.\ J.\  {\bf 391} (1992) L1.

  
\bibitem{us1}  H.~Falomir, J.~Gamboa, F.~Mendez and P.~Gondolo,
  Phys.\ Rev.\ D {\bf 96}, 083534 (2017).




\bibitem{us2} H.~Falomir, J.~Gamboa, F.~Mendez and P.~Gondolo, preprint 2018.



\bibitem{np}  V.~P.~Nair and A.~P.~Polychronakos,
  Phys.\ Lett.\ B {\bf 505}, 267 (2001)

  \bibitem{bellu} S.~Bellucci, A.~Nersessian and C.~Sochichiu,
  Phys.\ Lett.\ B {\bf 522}, 345 (2001)
  \bibitem{ccgm} J.~M.~Carmona, J.~L.~Cortes, J.~Gamboa and F.~Mendez,
  JHEP {\bf 0303}, 058 (2003).


\end{thebibliography}
\end{document}